 \documentclass[aps,prd,twocolumn,showpacs,preprintnumbers]{revtex4}

\usepackage{graphicx}
\usepackage{epsfig}
\usepackage{textcomp}
\usepackage{color}

\def\theta{\vartheta}
\def\nue{{\nu_e}}
\def\anue{{\bar\nu_e}}

\newcommand{\be}{\begin{equation}}
\newcommand{\ee}{\end{equation}}
\newcommand{\ba}{\begin{eqnarray}}
\newcommand{\ea}{\end{eqnarray}}
\newcommand{\lsim}   {\mathrel{\mathop{\kern 0pt \rlap
  {\raise.2ex\hbox{$<$}}}
  \lower.9ex\hbox{\kern-.190em $\sim$}}}
\newcommand{\gsim}   {\mathrel{\mathop{\kern 0pt \rlap
  {\raise.2ex\hbox{$>$}}}
  \lower.9ex\hbox{\kern-.190em $\sim$}}}

\begin{document}

\title{Testing Lorentz invariance with neutrino bursts from supernova neutronization }

\author{Sovan Chakraborty}
\affiliation{ II Institut f\"ur Theoretische Physik, Universit\"at Hamburg, Luruper Chaussee 149, 22761 Hamburg, Germany} 
\author{Alessandro Mirizzi}
\affiliation{ II Institut f\"ur Theoretische Physik, Universit\"at Hamburg, Luruper Chaussee 149, 22761 Hamburg, Germany} 
\author{G\"unter Sigl}
\affiliation{ II Institut f\"ur Theoretische Physik, Universit\"at Hamburg, Luruper Chaussee 149, 22761 Hamburg, Germany}

\begin{abstract}
 Quantum-gravity effects might generate Lorentz invariance violation by the interaction of energetic particles with the foamy  structure of the space-time. As a consequence, particles may not travel at the universal speed of light. We propose to constrain the Lorentz invariance violation for energetic neutrinos exploiting the $\nu_e$ neutronization burst from the next galactic supernova (SN). This prompt signal is expected to produce a sharp peak in the SN $\nu_e$ light curve with a duration of  $\sim 25$~ms. However, the presence of the energy-dependent Lorentz invariance violation would significantly spread out the time structure of this signal. We find that the detection of the SN $\nu_e$ burst  from a typical galactic explosion at $d=10$~kpc in a Mton-class water Cerenkov detector, would be sensitive to a quantum-gravity mass scale $M_{\rm QG} \sim 10^{12}$ GeV ($2 \times10^{5}$ GeV) for the linear (quadratic) energy dependence of the Lorentz invariance violation. These limits are valid for both super and subluminal neutrino velocity and are also independent of the neutrino mass hierarchy. 
 \end{abstract}


\maketitle

\section{Introduction}\label{Intro}
Lorentz invariance violation (LIV) arises in many approaches of quantum-gravity (QG) theories, suggesting Lorentz symmetry might be violated at very high energies (see  \cite{Liberati:2009pf} for a review).
In this context,  the space-time foam attributable to QG fluctuations might cause energetic
particles to propagate at speed $v$ different from the velocity of light $c$, which would be approached only by low-energy massless particles.
This LIV effect can be phenomenologically parametrized as
\begin{equation}
\frac{v}{c} = 1 \pm \left(\frac{E}{M_{QG}}\right)^n \,\ ,
\end{equation}
where $n=1,2$ denote linear or quadratic deviation for superluminal (+) 
or subluminal  (-) particles, respectively, occurring at a mass scale 
$M_{QG}$.  

Apart from photons \cite{Ellis:2005}, neutrinos from  astrophysical sources can also be useful to constrain
LIV effects. In case of $\sim {\mathcal O} (\rm{GeV})$ energy neutrinos from gamma ray bursts or active galactic nuclei~\cite{Jacob:2006gn} one expects to detect at most one or two such neutrinos. Therefore in such sources, the main attempt to constrain the LIV scale is through the energy dependent time of flight delay between the neutrinos and the corresponding $\gamma$ rays. Unfortunately till now, no high-energy $\nu$ from these astrophysical sources has been detected. However, this is not the case with core-collapse SNe as they are  guaranteed emitters of  MeV neutrinos, that  have been widely discussed in literature to  put bounds on LIV. In particular, from SN1987A data, exploiting the flight delay of a few hours of the $\gamma$ with respect to  $\nu$, a bound $(v-c)/c \lesssim 10^{-12}$ has  been obtained \cite{Stodolsky:1987vd}, corresponding to $M_{QG} \gtrsim 10^{9}$~GeV for $n=1$, or $M_{QG} \gtrsim 10^{3}$~GeV for $n=2$, taking $E\simeq 10$~MeV as typical SN $\nu$ energy.

Moreover, the expected duration of the SN1987A $\nu$ burst ($\sim 10$~s) was in agreement with the observed one, thus limiting LIV effects that would have broadened the SN $\nu$ time structure due to the energy dependency of the Lorentz violation. From the absence of an anomalous dispersion, it has been found  $M_{QG} \gtrsim 2\times 10^{10}$~GeV for the $n=1$ case, and $M_{QG} \gtrsim 4\times 10^{4}$~GeV for the $n=2$ case \cite{Ellis:2008fc}. These bounds are stronger than the previous  ones obtained from the SN $\nu$  time of flight because the observed time dispersion is a few tens of seconds, and a detailed statistical
analysis has been performed. Moreover, these limits are also much more stringent than the ones obtained from the $\nu$ time of flight measurements in the  long-baseline MINOS \cite{Adamson:2007zzb}, OPERA \cite{Adam:2011zb} and ICARUS experiments \cite{Antonello:2012hg}, giving $M_{QG} \gtrsim  10^{5}$~GeV for the $n=1$ case and  $M_{QG} \gtrsim  10^{2}$~GeV for the $n=2$  case.
 
\begin{figure}[!t]
\begin{center}  
\includegraphics[width=0.7\columnwidth,angle=270]{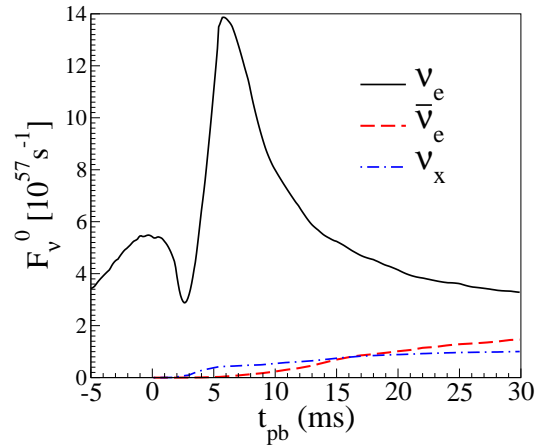}
\end{center}

\caption{\label{fig1}
Time evolution of the initial neutrino number flux
for  a 15 $M_{\odot}$ SN simulation from the Garching group, for $\nue$ (continuous curve), $\anue$ (dashed curve) and $\nu_x$ (dot-dashed curve).
}
\end{figure}

 One  expects that these bounds could be significantly improved  with the observation of a high-statistics signal from the next galactic SN. In this regard, it has been estimated that from the time structure of the $\nu$ signal from a SN at $d=10$~kpc observable in the Super-Kamiokande detector (with fiducial mass 22.5~kton), it would be possible to constrain $M_{QG} \gtrsim 2(4)\times 10^{11}$~GeV for subluminal  (superluminal) propagation in the $n=1$ case and $M_{QG} \gtrsim 2(4)\times 10^{5}$~GeV for subluminal (superluminal) propagation in the $n=2$ case \cite{Ellis:2008fc}. An even stronger bound (up to two orders of magnitude beyond previous estimates) could be obtained exploiting  variations in time on the scale of a few milliseconds found in multidimensional SN simulations \cite{Ellis:2011uk}. However, these features could probably be detected only for a very close-by SN (at $d \leq 2$~kpc) \cite{Lund:2012vm}. In the following, we propose to explore the effects of  LIV on the prompt SN $\nu_e$ neutronization burst. This signal is a common feature \cite{Kachelriess:2004ds} found in all sophisticated supernova simulations. Physically the newly formed SN shock disintegrate the dense iron core into free neutrons and protons. The electron rich environment behind the SN shock triggers rapid electron capture on these free protons producing a huge $\nu_e$ flux. Thus a large number of $\nu_e$'s are released when the shock breaks the neutrinosphere and the $\nu_e$'s escape freely. This process of deleptonization ``neutronizes'' the environment.  The $\bar\nu_e$ and muon and tau neutrino-(anti)neutrino (denoted with $\nu_x$) luminosities rise much more slowly compared to the $\nu_e$'s. These facts are very well depicted in Fig.~\ref{fig1}, where we plot the neutrino number flux of different flavors in the neutronization phase for a 15$M_{\odot}$ SN simulation from the Garching group \cite{Kachelriess:2004ds}. The signature of the neutronization peak is nearly independent from the details of SN models, like electron capture rates, nuclear equation of state and the progenitor mass. Indeed, it can be considered as a ``standard neutrino candle'' in the simulations of core-collapse supernovae.  In fact, probing the time-spectra of the neutronization peak one can distinguish the case of inverted neutrino mass hierarchy ($\Delta m_{\rm atm}^2=m_{2,3}^{2}-m_{1}^{2}<0$) from the normal mass hierarchy ($\Delta m_{\rm atm}^2>0$) \cite{Kachelriess:2004ds}, where the peak is suppressed for a value of  a 1-3 mixing angle $\theta_{13}$ as large as the one currently measured by reactor experiments \cite{An:2012eh,Ahn:2012nd}. The LIV inducing an energy dependence in the $\nu$ velocity will broaden the time structure to the $\nu_e$ burst as neutrinos with different energies will undergo different time delays.\\ 
In the following we characterize this effect. In Sec.~\ref{flavour}, we describe the SN neutrino flavor conversions  during the $\nu_e$ burst phase. In Sec.~\ref{detection} we recall the calculation  of the SN neutrino events rate  in a Mton class water Cherenkov detector. In Sec.~\ref{results} we present the effects of the LIV on the $\nu$ signal during the neutronization burst. Finally, we summarize our results in Sec.~\ref{conclusion}.
\vspace{-0.4cm}

\section{Neutrino flavor conversions}
\label{flavour}
The SN neutrino signal observed at Earth  is processed by flavor conversions during the propagation in the stellar
envelope.
Since negligible
$\bar\nu_e$ and $\nu_x$ fluxes are emitted during the neutronization phase, self-induced flavor
conversions are absent \cite{Hannestad:2006nj}.   Thus, only Mikheyev-Smirnov-Wolfenstein flavor transitions occur while the neutrinos propagate through the stellar envelope \cite{Matt}.
Given 
 the discovery of large $\theta_{13}$ \cite{An:2012eh,Ahn:2012nd}  the SN  $\nu$ flux at Earth is straightforward to calculate.
Namely,
the $\nu_e$ and $\bar\nu_e$ fluxes 
 in  normal hierarchy (NH) are given by \cite{Dighe:1999bi,Chakraboty:2010sz}  
\begin{eqnarray}
F_{\nue} & = & F^{0}_{\nu_x}\, \\
F_{\anue} & = & \cos^2 \theta_{12} (F^{0}_{\anue}-F^{0}_{\nu_x}) + F^{0}_{{\nu}_x}.
\label{eq:nh}
\end{eqnarray}
while in
inverted hierarchy (IH) one gets
\begin{eqnarray}
F_{\nue} & = & \sin^2 \theta_{12} (F^{0}_{\nue}-F^{0}_{\nu_x}) + F^{0}_{\nu_x}\, \\
F_{\anue} & = &  F^{0}_{{\nu}_x}.
\label{eq:ih}
\end{eqnarray}
Here $\theta_{12}$ is the 1-2 mixing angle, with $\sin^2 \theta_{12}  \simeq 0.31$~\cite{Fogli:2012ua}, and $F^{0}_{\nu_{\alpha}}$ is the initial flux for the $\alpha$'th
flavor. The fluxes of the other flavors at Earth can be found from flavor conservation,
 $F_{\nue}+ 2 F_{\nu_x}= F^{0}_{\nue} + 2 F^{0}_{\nu_x}$ and $F_{\anue}+ 2 F_{\bar\nu_x}= F^{0}_{\anue} + 2 F^{0}_{\bar\nu_x}$.
\vspace{-0.4cm}
\section{Neutrino event rate}
\label{detection}
\begin{figure*}[!]
\epsfig{file=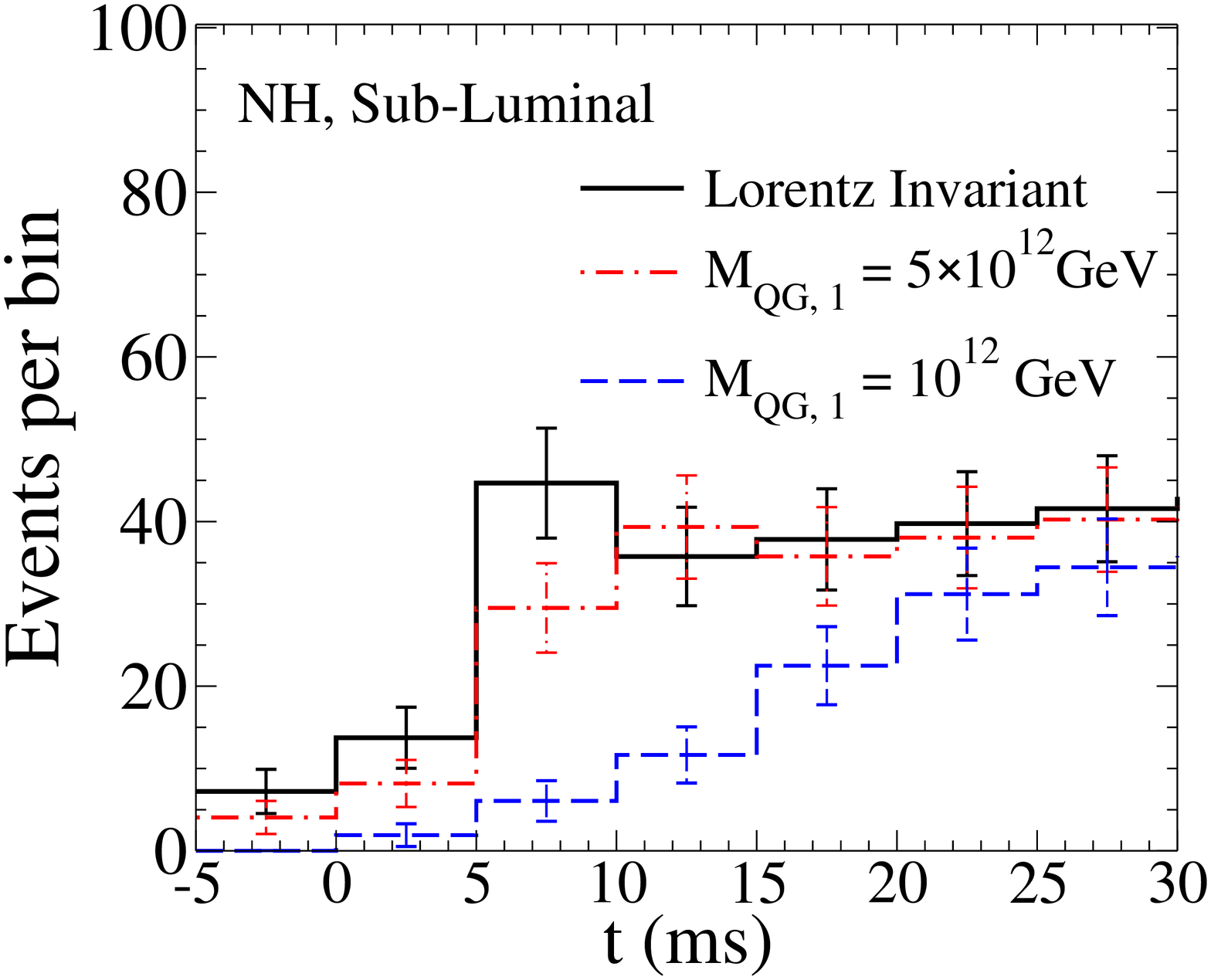,width=0.35\hsize,angle=270}
\epsfig{file=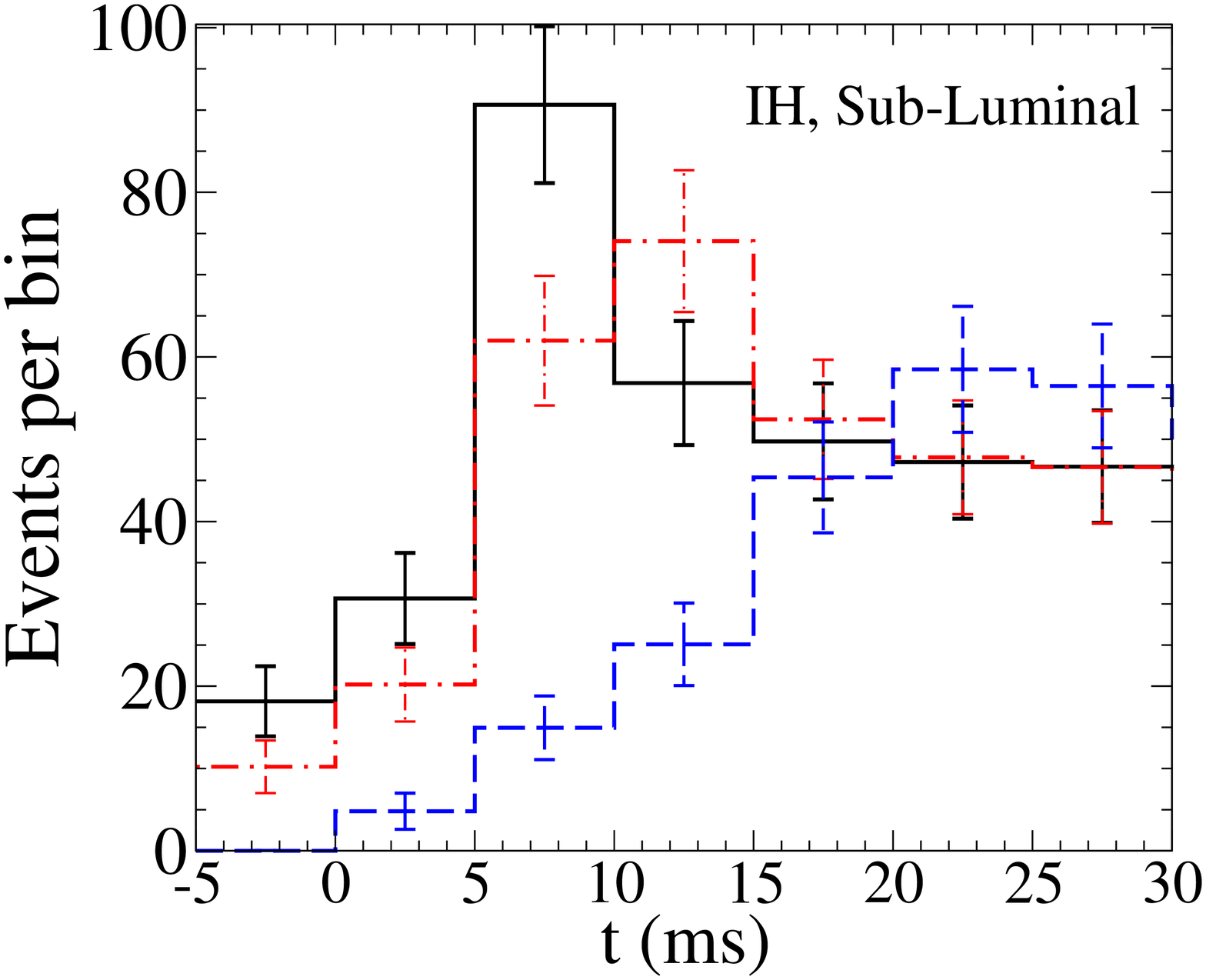,width=0.35\hsize,angle=270}
\epsfig{file=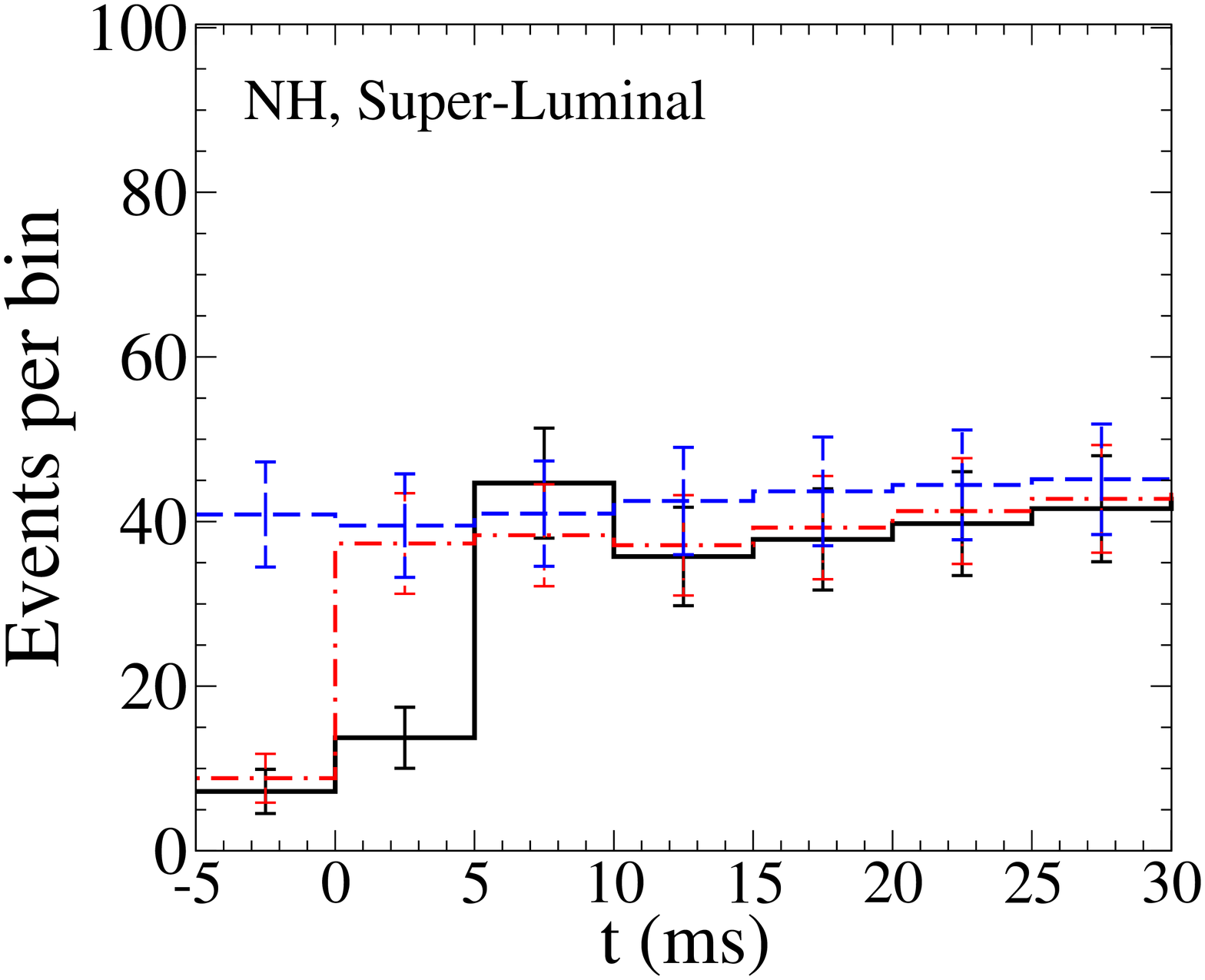,width=0.35\hsize,angle=270}
\epsfig{file=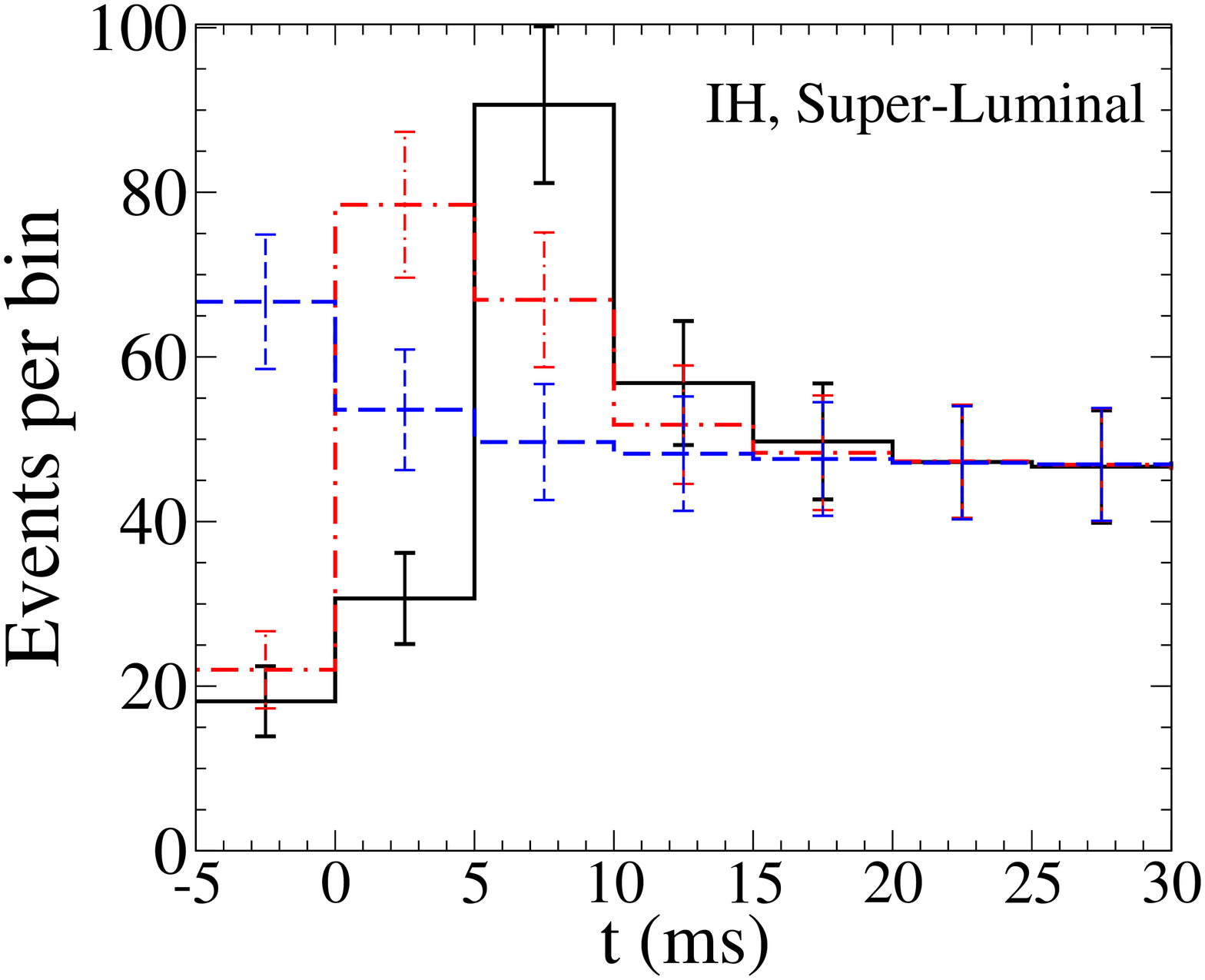,width=0.35\hsize,angle=270}
\caption{\label{fig2}
Neutronization events rate per time bin in a 740 kton water Cerenkov detector 
for a SN at $d=10$~kpc. The upper two panels describe subluminal LIV, whereas the lower panels are for the superluminal scenario. The right panels are for NH and the left ones for IH. The Lorentz invariant (LI) cases are shown in continuous lines. For LIV we consider the linear energy dependence, i.e., $n=1$. The  dot-dashed lines describe LIV for  $M_{QG}= 5 \times10^{12}$~GeV and the dashed lines describe LIV for $M_{QG}= 10^{12}$~GeV. The error bars reflect the number of expected neutrino events in each time bin.
}
\end{figure*}
A clear detection of the $\nu_e$ neutronization burst requires large detectors with a sensitivity to the electron neutrino flavors. In particular, a liquid argon time projection chamber with a mass of ${\mathcal O}(100)$~kton would be able to distinguish this signal via charged-current reactions on Ar nuclei \cite{Gil-Botella:2003sz}. Moreover, also Mton class water Cherenkov detectors could get a signature for the $\nu_e$ neutronization burst,  through the elastic scattering of $\nu_e$ on electrons ($\nu_e+e^-\rightarrow \nu_e+e^-$). In the following, for definitiveness we focus on a future water Cherenkov detector with a fiducial mass of 740 kton \cite{Abe:2011ts}, like the Japanese Hyper-Kamiokande experiment. For our calculations,  we take as a benchmark the results of the 15 $M_{\odot}$ Garching simulations shown in Fig.~\ref{fig1} to characterize the original time-dependent SN neutrino fluxes [$F_{\nu} = F_{\nu}(E,t_i)$] during the early neutrino emission phase. Here $E$ is the neutrino energy and $t_i$ is the emission time at the source. 

The arrival time of a neutrino of energy $E$ in the detector at Earth is $t_\oplus = t_i + d/c + \bigtriangleup t $, where $d$ is the distance of the source. The energy dependent time delay owing to QG is given by 
\begin{equation}
 \bigtriangleup t = \pm \frac{d}{c} \left(\frac{E}{M_{QG}}\right)^{n} \,\ ,
 \end{equation}
where the (+) sign is for the subluminal and (-) for the superluminal case. The constant term $d/c$ in $\bigtriangleup t$ is dropped for convenience. Then the event rates of neutrinos at the detector from a SN at distance $d$, is given by

\vspace{-0.1cm}
\begin{eqnarray*}
\frac{dN}{dt}(t_\oplus)
& = & \frac{n_{T}}{4 \pi d^{2}} {\int}\,dE\,\sigma (E)\,{\int\,dt_{i}\,F_{\nu}(E,t_{i})\,\delta(t_\oplus - t_i - \bigtriangleup t)}\, \\
& = & \frac{n_{T}}{4 \pi d^{2}} {\int}\,dE\,\sigma (E)\,F_{\nu}(E,t_\oplus  - \bigtriangleup t).\, \\
\label{eq:evntrate}
\end{eqnarray*}

Here $n_{T}$ is the number of target nucleons, $\sigma (E)$ the elastic-scattering cross section for a neutrino energy $E$ on the target \cite{Kachelriess:2004ds}. In the following we will assume a typical SN at $d=10$~kpc. Our calculation of the events rate closely follows the one presented in \cite{Kachelriess:2004ds} to which we refer the interested reader for further details.

\vspace{-0.4cm}
\section{Results}\label{results}

In Fig.~\ref{fig2} we show the event rate expected during the neutronization burst for different scenarios. The left panels refer to the NH cases, while the right ones refer to IH. Moreover, the upper panels
show the subluminal scenarios and the lower ones the superluminal ones. In the absence of Lorentz violation (continuous curves) the difference between the time structure of the events for the two different mass hierarchies reconfirms the  results presented in \cite{Kachelriess:2004ds}. Since in the NH case the signal  is dominated by $F^{0}_{\nu_x}$ [see Eq. (\ref{eq:nh})] the peak is suppressed. Instead, for the IH, the flux is dominated by $F^{0}_{\nue}$ [see Eq. (\ref{eq:ih})] and the signal has a peaked time spectrum. 
\begin{figure*}[!]
\epsfig{file=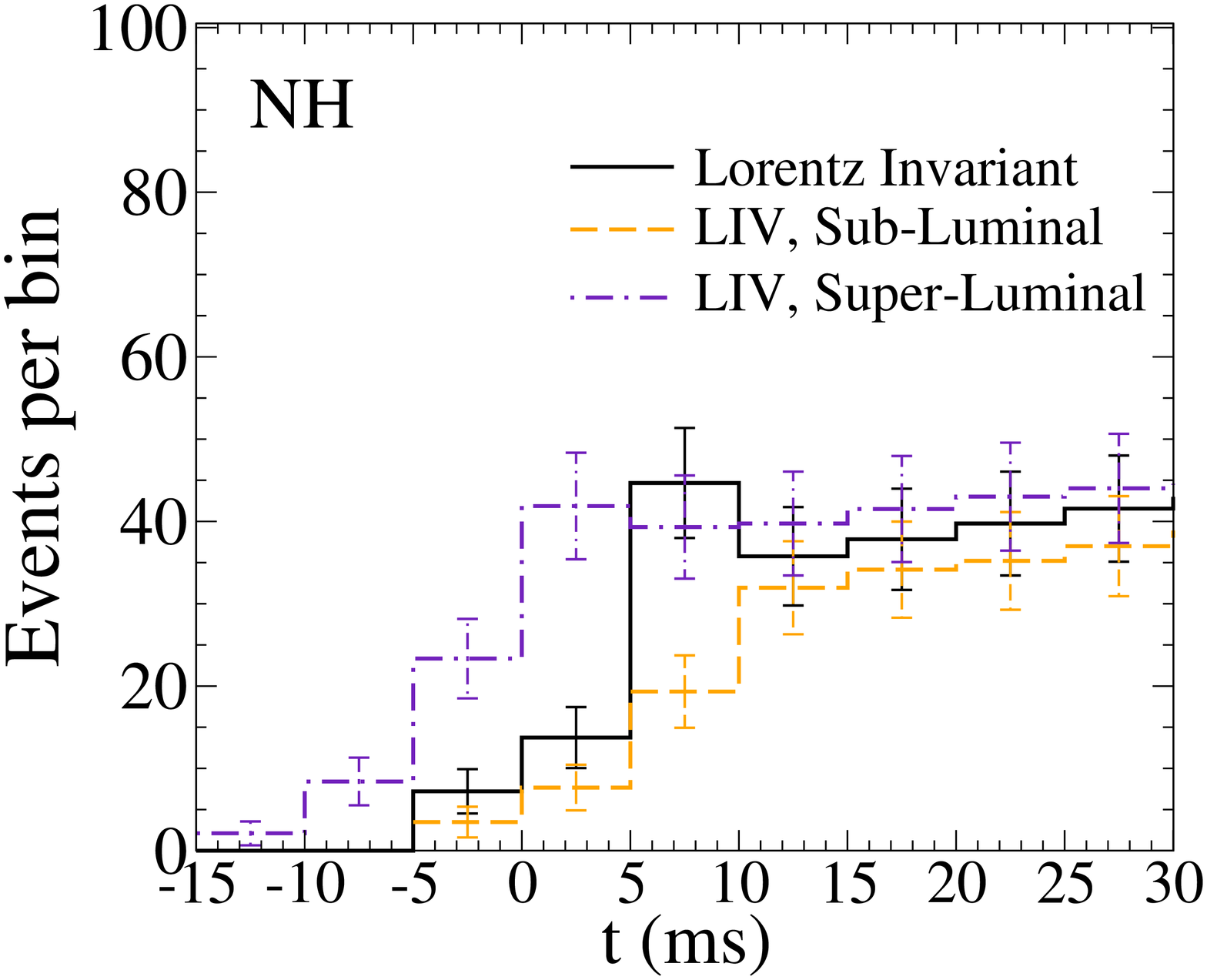,width=0.34\hsize,angle=270}
\epsfig{file=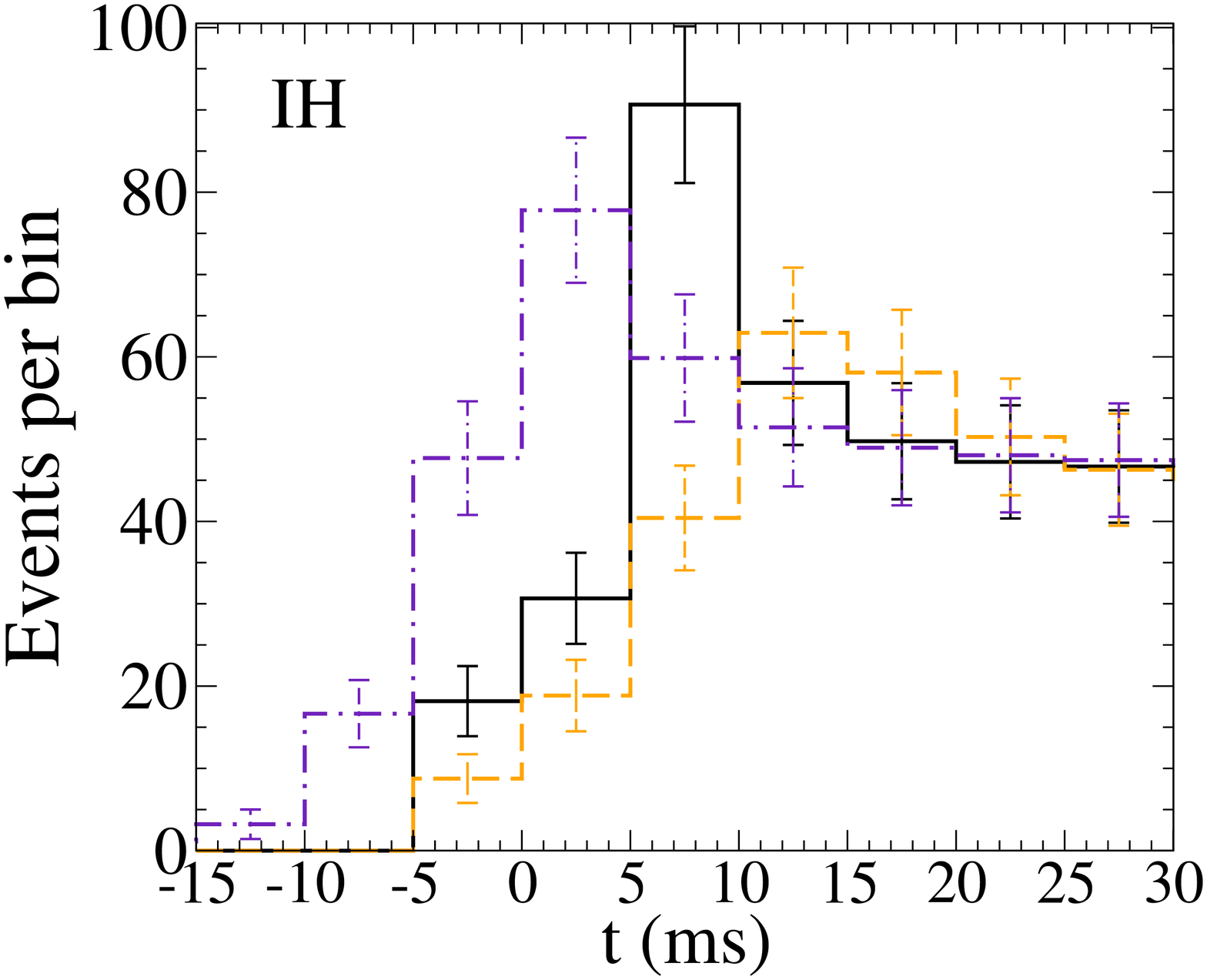,width=0.34\hsize,angle=270}
\caption{\label{fig3}
Neutronization event rate per time bin in a 740 kton water Cherenkov detector for a SN at $d=10$~kpc. Here the energy dependence of LIV  is quadratic, i.e $n=2$. The left panel refers to the NH case, while the right panel refers to IH. The cases without LIV are represented by the continuous curves. In the cases with LIV we take as representative QG mass scale  $M_{QG}= 2 \times10^{5}$~GeV. The subluminal cases are in dashed curves, while the superluminal are in dot-dashed  ones.}
\end{figure*}
The standard scenario significantly changes once the LIV effects are included. We show examples with the linear energy dependence case, i.e., $n=1$ with $M_{QG}= 10^{12}$~GeV (dashed curves) and $5 \times 10^{12}$~GeV (dot-dashed curves). LIV tends to flatten and spread out the  original time structure in the signal. In particular, in the subluminal case for $M_{QG}= 5 \times 10^{12}$~GeV the strength of the peak is reduced and shifted to later times (by $\sim 5$~ms), while for  $M_{QG}=10^{12}$~GeV the peak is completely washed out and the signal presents a shape that monotonically rises with time. In the superluminal cases, for $M_{QG}= 5 \times 10^{12}$~GeV the peak is shifted to earlier times in both hierarchies (by $\sim 5$~ms). For NH with $M_{QG}= 10^{12}$~GeV the time spectrum becomes flat, whereas in IH it monotonically decreases.

We also considered the case of quadratic energy dependence of LIV in Fig.~\ref{fig3}.  Absence of  LIV features would constrain  $M_{QG} \gtrsim2\times10^{5}$ GeV in both superluminal and subluminal cases. 

\vspace{0.0 cm}
\section{Summary}\label{conclusion}
\vspace{-0.2 cm}
We have studied the effects of Lorentz invariance violation in the neutrino sector on the SN neutronization burst. The analysis shows that the Lorentz invariance violation would produce a strong suppression of the expected neutronization peak in both the superluminal and subluminal cases. We find a sensitivity to the quantum-gravity mass scale of $M_{\rm QG} \sim 10^{12}$ GeV ($2 \times10^{5}$ GeV) for the linear (quadratic) energy dependence of LIV in both the neutrino mass hierarchies.

This method would give a clean bound on quantum-gravity mass scale, better and simpler than the present limits from SN neutrinos. Indeed these potential limits come from the different shapes of the neutronization peak, which is a rather model independent feature of SN simulations.

\vspace{-0.6cm}
\section{Acknowledgments}\label{ack}
\vspace{-0.4cm}
We thank Thomas Janka for providing us with SN neutrino data from their simulations. We also thank Mark Vagins for useful discussion on Hyper-Kamiokande. This work was supported by the German Science Foundation (DFG) within the Collaborative Research Center 676 ``Particles, Strings and the Early Universe,'' by the ``Helmholtz Alliance for Astroparticle Physics (HAP)'' funded by the Initiative and Networking Fund of the
Helmholtz Association, and by the State of Hamburg, through the Collaborative Research program (LEXI) ``Connecting Particles with the Cosmos.''


\end{document}